\documentclass[journal=ancac3,manuscript=article]{achemso}

\usepackage{amsmath}
\usepackage{graphicx}
\usepackage{natbib}

\newcommand{\onlinecite}[1]{\hspace{-1 ex} \nocite{#1}\citenum{#1}}

\author{Alejandro A. Pacheco Sanjuan}
\affiliation{Department of Physics. University of Arkansas. Fayetteville AR 72701. USA}
\alsoaffiliation{Department of Mechanical Engineering. Universidad del Norte. Barranquilla, Colombia}
\author{Mehrshad Mehboudi}
\affiliation{Department of Physics. University of Arkansas. Fayetteville AR 72701. USA}
\author{Edmund O. Harriss}
\affiliation{Department of Mathematical Sciences. University of Arkansas. Fayetteville AR 72701. USA}
\author{Humberto Terrones}
\affiliation{Department of Physics and Center for 2-Dimensional and Layered Materials. The Pennsylvania State University. University Park PA 16802. USA}
\author{Salvador Barraza-Lopez}
\affiliation{Department of Physics. University of Arkansas. Fayetteville AR 72701. USA}
\email{sbarraza@uark.edu}
\title{Quantitative Chemistry and the Discrete Geometry of Conformal Atom-Thin Crystals}


\begin{document}


\begin{abstract}
 When flat or on a firm mechanical substrate, the atomic composition and atomistic structure of two-dimensional crystals dictate their chemical, electronic, optical, and mechanical properties. These properties change when the two-dimensional and ideal crystal structure evolves into arbitrary shapes, providing a direct and dramatic link among geometry and material properties due to the larger structural flexibility when compared to bulk three-dimensional materials. We describe methods to understand the local geometrical information of two-dimensional conformal crystals quantitatively and directly from atomic positions, even in the presence of atomistic  defects. We then discuss direct relations among the discrete geometry and chemically-relevant quantities --mean-bond-lengths, hybridization angles and $\sigma-\pi$ hybridization. These concepts are illustrated for carbon-based materials and ionic crystals. The piramidalization angle turns out to be linearly proportional to the mean curvature for relevant crystalline configurations. Discrete geometry provides direct quantitative information on the potential chemistry of conformal two-dimensional crystals.
\end{abstract}

\noindent{}{\bf Keywords:}\\
\noindent{}Conformal geometry, two-dimensional crystals, pyramidalization angle, discrete geometry.\\


 The time could not be more ripe to reinvigorate the long-standing discussion \cite{NewGeometriesBook,DresselhausBook,reviewDresselhausTerrones} as to how the properties of elastic two-dimensional (2D) conformal elastic crystals \cite{NovoPNAS} are affected by geometry \cite{Kuhnel,Math1}. 2D crystalline membranes adapt to the shape of ({\em i.e.}, they conform to) harder surfaces\cite{PRLMorozov2006,Berry,PrincessPea}, develop ripples when freestanding \cite{Nature2007,Biro,Shenoy,Liu,HWang,curtains,us,usSSC,Fasolino1,Zakharenko,Katsnelson2,PRB2010}, and can be deformed into arbitrary elastic regimes beyond harmonic elasticity theory \cite{Hone1,Hone2,Crommie,stroscio,usold,nanoletters2stms}. The field is taking off spectacularly with new materials coming into play and a host of ideas from graphene physics --such as strain engineering  \cite{Ando2002,Pereira1,GuineaNatPhys2010,Vozmediano,deJuanPRB,Dejuan2011,deJuanPRL2012,deJuan2013,Kitt2013,Peeters3,Peeters4,PeetersNew1,PeetersNew2,Naumis1,Naumis2,Manes,Crommie,usold,stroscio,deheer2,us,usSSC,usgeo}-- being applied to other 2D materials with more diverse electronic, valley and spin properties \cite{BNsinglelayer,Grigorieva2013,Henning1,Henning2,strainGuinea2013,review2013}. At the present moment the portfolio of 2D crystals includes graphene \cite{Wallace,Mclure,Semenoff,DiVicenzoMele,Geim1,kim,RMP}, hexagonal boron nitride, transition-metal dichalcogenides, some oxides \cite{NovoPNAS,BNsinglelayer,other1,other2,other3,other5,other6,o8}, silicene and germanene \cite{silicene}, and new semiconductor compounds \cite{Henning1,Henning2}. Their material properties will depend on shape.

Bulk three-dimensional materials cannot exceed a few-percent elastic strain. In contrast, freestanding 2D crystals can sustain much larger metric increases and strain into the tens of percent (with or without accompanying curvature); they can be driven away from the harmonic elastic regime with ease \cite{Hone1,Hone2,mos21,mos22,mos23,mos25,mos26}.
 Elastic two-dimensional membranes \cite{Bookmembranes1,Bookmembranes2} can assume the shape of arbitrary two-dimensional manifolds \cite{Seung88,NewGeometriesBook,t1,t2,t3,t4}. Flexible 2D membranes relieve mechanical in-plane strain originating from defects by buckling out-of-plane, hence trading stretching by bending \cite{Seung88}. The geometry of atomistic membranes can also be determined by the presence of atomistic (topological) defects or lack thereof \cite{Seung88,WalesPRB2009}. Atomistic defects also induce radical-behavior by localizing electronic orbitals, altering optically-available electronic transitions, and can also induce a large curvature and strain \cite{DresselhausBook,NewGeometriesBook}.

 When thinking of crystalline membranes, geometry tends to be linked to curvature alone \cite{NewGeometriesBook}, but when distances among atoms change --even under in-plane strain-- the metric changes as well. Since curvature and metric both determine the local geometry, conformations with and without in-plane strain are geometrically distinct already. Furthermore, mechanical strain and metric are directly related --and may even be used interchangeably. While mechanical strain may be modeled by a host of mechanical theories, each with their own assumptions ({\em i.e.}, continuum mechanics through the Cauchy-Born rule\cite{Landau} {\em versus} atomistic mechanics with interatomic potentials \cite{LAMMPS}, all the way to full {\em ab-initio} based molecular dynamics \cite{parinello}), one can bypass mechanistic details and couple the relevant optical, electronic and chemical theories  to changes in interatomic distances directly. This concept was pioneered for quantum dots under strain since the late eighties \cite{zunger} and it acquires a renewed relevance for 2D crystals, because the structural analysis directly originates from finite atomic displacements, instead of being mediated by a continuum. By progression of thought, these concepts arguably give atomistic (non-continuum) approaches to geometry a special relevance too. Those atomistic approaches to geometry are based on meshes\cite{Math1,chinese}, but the identification of an abstract mesh with the atomic lattice of 2D crystals is a direct one \cite{usgeo}, and a number of teams are already complementing continuum mechanical \cite{ariza1,ariza2} and electronic structure \cite{us,usSSC,usgeo,PeetersNew2,PereiraNL2013} descriptions of crystalline membranes with some decidedly discrete approaches.

We argue that the mathematical tool for geometrical analysis is not just an object to process, filter, and/or fit physical and chemical information. Instead, the geometrical language can in fact lead and motivate the scientific discussion. This observation is particularly acute regarding Chemistry. Within continuum geometry/elasticity theory, a 2D membrane is a chemically inert atom-less medium, and no discussion of chemical reactivity (or aromaticity for carbon-based materials) has ever originated from within continuum elasticity as far as we know.

 On the other hand, consider as an example benzene-like carbon materials (such as graphene), where {\em aromaticity} is a structural property \cite{Wu}. Infinite ideal 2D graphene has three equivalent aromatic Clar structures that lead to a constant mean bond length (MBL) (Figure 1(a)) \cite{Wu,randic,mauri,r39,r40,r47,r48,r49,r51,r56,r57,r59}: This is identical to say that the metric (Figure 1(c)) is constant in such a system, thus suggesting an intimate connection among aromatic behavior and geometry. Now that the link among the aromaticity of carbon rings and the metric has been exposed, the next interesting and related question from a chemical perspective is: What is the reactivity/aromaticity of strain-engineered graphene \cite{Pereira1,Vozmediano,GuineaNatPhys2010,Dejuan2011}? As far as we know, this question is yet to be addressed; the issue of chemical reactivity under non-planar structural conformations will certainly become more relevant as the Science of conformal 2D crystals continues to be explored in years to come.

Now, relevant for orbital hybridization \cite{Manes}, spin-orbit coupling \cite{Ando2000,Huertas-Hernando} and chemical reactivity, the {\em pyramidalization angle}\cite{Haddon} $\alpha_{pyr}$ relates to the angular deviation among the local normal and bond vectors (Figure 1(b)). The fact that curvature (obtained from points in Figure 1(c)) and $\alpha_{pyr}$ could be related may be expected intuitively, but their explicit dependence has not been discussed so far. Their explicit relation will be one of the main results of this paper.

\begin{figure}[bt]
\includegraphics[width=0.65\textwidth]{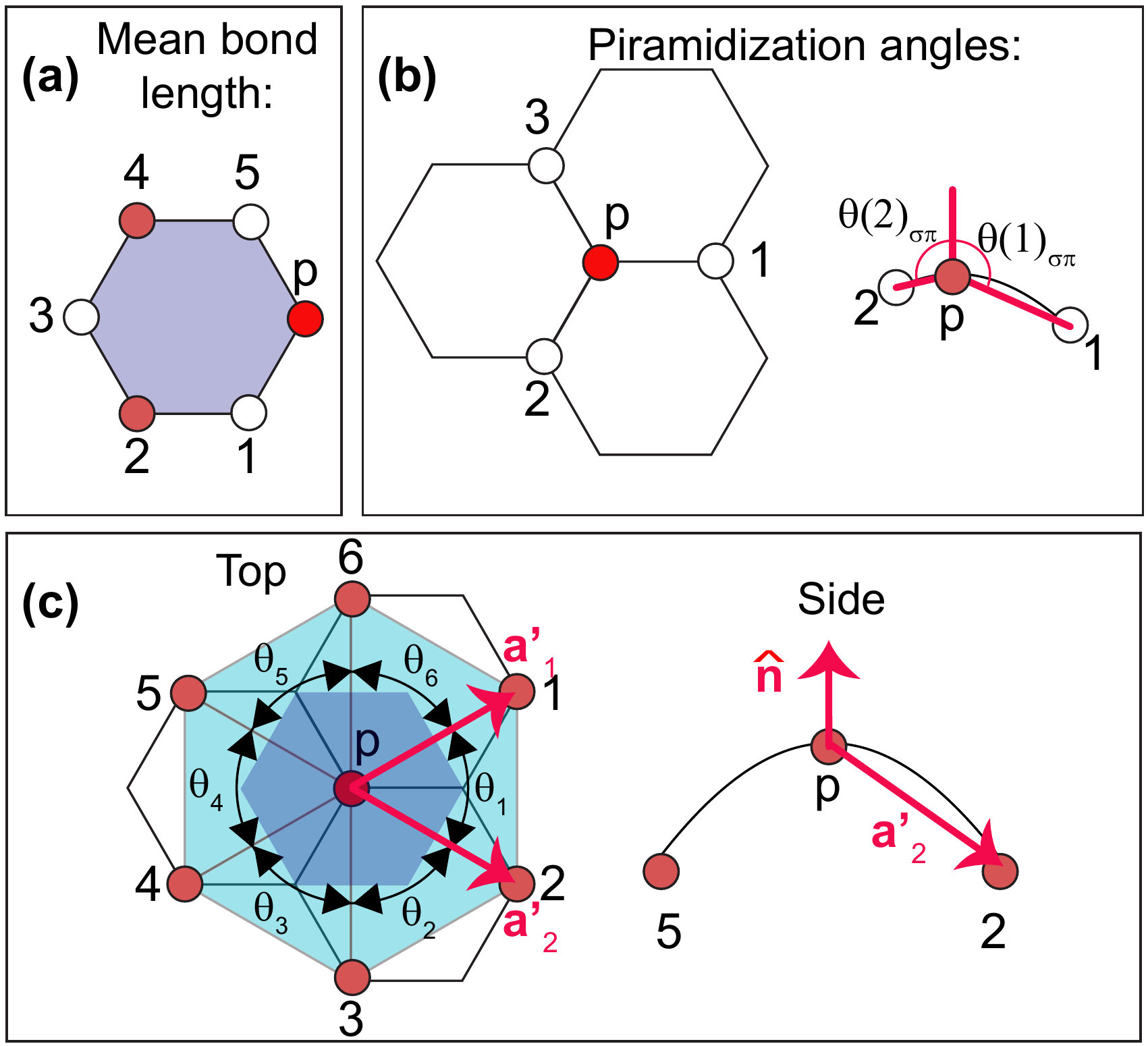}
\caption{Atoms required to determine local chemical measures and the geometrical parameters in Eqs.~(1-3) and (5-7) for 2D crystals with hexagonal structure and no bucking. (a) Atoms required to determine mean-bond lengths on a hexagon. (b) Four atoms needed to determine pyramidalization angles. (c) Local lattice displacements $\mathbf{a}_1'$ and $\mathbf{a}_2'$, internal angles $\theta_i$ to edges $\mathbf{e}_i$ and $\mathbf{e}_{i+1}$, and darker shaded hexagon of area $Ap$ are shown; side view shows the normal vector $\hat{\mathbf{n}}$.}
\end{figure}

 Hence,  we argue by example that geometry originating from atoms is a powerful concept leading to new scientific perspectives, new intuition, and to precise yet intuitive links to potential chemical and physical phenomena. In addition, although the systems discussed here explicitly are carbon-based, the general ideas are expected to apply for many 2D crystals, because the discussion here is of a general character as it pertains to geometry. Most of the phenomena at the single-electron level has to be explainable by (i) chemical composition, and (ii) electron orbital symmetries \cite{mintwire2,gengeo1,gengeo2,gengeo3}, (iii) thermodynamics \cite{Wales2,Wales3}, and other dynamical information at the relevant electron energies. Geometry is a control handle for these properties.

 The geometrical analysis applies unchanged to graphene, nanotubes  \cite{Ijima,DresselhausBook,nanotubes1,nanotubes2,nanotubes3,nanotubes5,nanotubes6,nanotubes7,nanotubes9,nanotubes11,nanotubes12,nanotubes14,nanotubes15,mintwire1,mintwire2,mintwire3,mintwire4,mintwire5}, single-layer hexagonal Boron Nitride \cite{BNsinglelayer}, and novel crystals with non-buckled hexagonal structure as those discussed by Hennig recently \cite{Henning1,Henning2}. The formalism is then extended by means of approximations to study arbitrary two-dimensional nanomaterials with more general shapes, such as fullerenes \cite{Kroto,fullerenes1,fullerenes2,fullerenes3,fullerenes5,fullerenes6}, Schwarzites \cite{H1}, ionic crystals \cite{Wales1,WalesPRB2009,WalesPRL2013,monica}, and other hexagonal systems with atomistic defects. Thicker conformal systems (such as single-layer transition metal dichalcogenides, or 2D systems with other than hexagonal lattices) could be later explored along similar lines.

\section{Chemical measures and structure}

 Three measures originate from carbon-based materials.

(1) Mean Bond Length (MBL): Aromaticity is not a directly measurable property and hence it cannot be defined unambiguously. Yet, the structure accommodating the maximum number of Clar sextets best represents chemical and physical properties, and Clar sextet migration increases chemical reactivity. How aromatic is conformal graphene \cite{Fasolino1}? Can rippling be explained in terms of the creation of ``aromatic domains''? The mean bond length (MBL) is defined as follows (Figure 1(a)) \cite{Wu,r49}:
\begin{equation}
MBL=\bar{a}_{CC}=\frac{1}{6}\sum_{i=1}^6 {a}_{CC,i},
\end{equation}
where ${a}_{CC,i}$ are bond lengths on a closed loop as depicted in Figure 1(a). According to Ref.~\onlinecite{r49}, $MBL$ is a reliable tool for analysis of large aromatic systems. To reach this conclusion, they compared their results with the pseudo-$\pi$ method \cite{FowlerSteiner} in the context of the six-center bond index (SCI) analysis on graphene nanoribbons.

(2) The angle $\theta_{\sigma\pi}$ between the bonds and the normal vector $\hat{\mathbf{n}}$ at atom $p$ has a single value $\theta_{\sigma\pi}$ under a spherical geometry \cite{Haddon} and $\theta_{\sigma\pi}=\pi/2$ on a flat surface. $\theta_{\sigma\pi}$ can be generalized for arbitrary geometries as an average:
\begin{equation}
\bar{\theta}_{\sigma\pi}\equiv\frac{1}{3}\sum_{i=1}^3\theta(i)_{\sigma\pi},
\end{equation}\label{eq:pa}
with $\theta(i)_{\sigma\pi}$ the angle among a bond vector and the local normal (Figure 1(b)). Equation (2) takes its usual form for fullerenes, where $\theta(i)_{\sigma\pi}=\theta_{\sigma\pi}$ for all bonds \cite{Haddon}. The pyramidalization angle $\alpha_{pyr}$ is defined by Haddon \cite{Haddon} as follows:
\begin{equation}
\alpha_{pyr}=\bar{\theta}_{\sigma\pi}-\pi/2.
\end{equation}

(3) $\sigma-\pi$ orbital hybridization: In the presence of curvature, the hopping parameter $t$ between nearest neighbors (1 and 2) for materials with $s$ and $p$-electrons is modified to include hybridization among $\sigma$ and $\pi$ electrons, as follows \cite{Manes}:
\begin{equation}
t\to (\hat{n}_1\cdot\hat{r})(\hat{n}_2\cdot\hat{r})(t_{\sigma}+\delta t_{\sigma})+[\hat{n}_1-(\hat{n}_1\cdot\hat{r})\hat{r}]\cdot [\hat{n}_2-(\hat{n}_2\cdot\hat{r})\hat{r}](t+\delta t).
\end{equation}
$\hat{n}_1$ and $\hat{n}_2$ are local normals for atoms 1 and 2, $\mathbf{r}$ is the vector joining them, and $\hat{r}=\mathbf{r}/r$.
For example, for a planar 2D crystal $\hat{n}_1$ and $\hat{n}_2$ are orthogonal to $\hat{r}$, and $t\to t+\delta t$ ({\em i.e.}, there is no $\sigma-\pi$ hybridization), as expected.  The effect of $\sigma-\pi$ electron coupling tends to be small and it is some times neglected but it is important to visualize the magnitude of the coupling coefficient $(\hat{n}_1\cdot\hat{r})(\hat{n}_2\cdot\hat{r})$ in relevant atomistic conformations.

 We next discuss geometry of atomic crystals in a comprehensive way.

\section{The local geometry of two-dimensional elastic membranes, from atomic positions}

 The local geometry of two-dimensional surfaces is determined from their metric ($g$) and curvature ($k$) in terms of four invariants that indicate how the two-dimensional manifold stretches and curves along two directions with respect to a reference non-deformed configuration. Given two in-plane vector fields $\mathbf{g}_1$ and $\mathbf{g}_2$, a well-defined metric  $g_{\alpha\beta}\equiv\mathbf{g}_{\alpha}\cdot \mathbf{g}_{\beta}$ must be symmetric ($g_{\alpha\beta}=g_{\beta\alpha}$) and positive definite ($g_{\alpha\alpha}>0$) \cite{Lee} ($\alpha,$ $\beta=1,2$). Then, the four invariant geometrical measures are: (i) $\det(g)=|\mathbf{g}_1|^2|\mathbf{g}_2|^2-|\mathbf{g}_1\cdot \mathbf{g}_2|^2$, $\text{Tr}(g)=\sum_{\alpha=1}^2|\mathbf{g}_{\alpha}|^2$, the Gauss curvature $K\equiv\det(k)/\det(g)$, and $H\equiv\sum_{\alpha=1}^2k_{\alpha\alpha}/2g_{\alpha\alpha}$, the mean curvature. (In a continuum approach, the curvature is derived from the metric tensor as follows: $k_{\alpha\beta}=\hat{\mathbf{n}}\cdot  \frac{\partial\mathbf{g}_{\alpha}}{\partial {\xi^{\beta}}}$ with $\xi^{\beta}$ an in-plane coordinate. See, {\em e.g.}, Ref.~\onlinecite{M4}.)

Two-dimensional conformal crystals can be studied with non-traditional tools that include discrete geometry, discrete differential geometry (DDG), and discrete exterior calculus. Classical, {\em Riemannian} differential geometry studies the properties of smooth, continuum objects, and {\em discrete} geometry studies geometrical shapes made of polyhedra. DDG, in turn, seeks discrete equivalents of notions and methods of continuous and discrete geometry, and applies to conformal 2D crystals  transparently. In fact, many concepts of smooth geometry are limiting procedures of discretizations. The main points for DDG are that interatomic distances determine the discretization limit, and that the continuum limit is not granted on atomistic meshes. As seen in the next subsection, both curvatures can be expressed within the framework of DDG.

\section{Results}
\subsection{Basic framework}

Discrete geometry is expressed from atomic positions, and the discrete metric is defined from lattice displacements $\mathbf{a}_{\alpha}'$ \cite{us,usSSC,usgeo} on the deformed lattice (Figure 1(c)):
\begin{equation}\label{eq:metric}
g_{\alpha\beta}=\mathbf{a}_{\alpha}'\cdot \mathbf{a}_{\beta}'
\end{equation}

 The Gauss curvature ($K$) originates from the {\em angle defect} $\sum_{i=1}^6\theta_i$ (Figure 1(c)) as follows \cite{Math1,Math2,chinese}:
\begin{equation}\label{eq:DGB}
K=(2\pi-\sum_{i=1}^6\theta_i)/A_p,
\end{equation}
where $\theta_i$ ($i=1,...,6$) are inner angles among vertices and the {\em Voronoi tessellation} [darker blue in Fig.~1(c)] encloses the area $A_p$ of a unit cell. Consistent with the existence of two atoms on the unit cell of 2D crystals with hexagonal symmetry\cite{Geim1,BNsinglelayer,Henning1}, the points $p$ and 1-6 on Fig.~1(c) belong to the same sublattice. On a flat surface, for example, the angle defect adds up to $2\pi$ making $K=0$, as expected.

 Still relying on Figure 1(c) for geometrical guidance, the mean curvature $H$ averages the relative orientations of edges and normal vectors along a closed path, and it is a signed quantity ({\em i.e.}, it is sensitive to the side of the 2D surface through its projection onto the local normal $\hat{\mathbf{n}}$):
\begin{equation}\label{eq:Hdiscrete}
H=\frac{\sum_{i=1}^6 \mathbf{e}_i\times(\boldsymbol{\nu}_{i,i+1}-\boldsymbol{\nu}_{i-1,i})}{4A_p}\cdot \hat{\mathbf{n}}.
\end{equation}
Here, $\mathbf{v}_i$ is the position of atom $i$ on sublattice $A$, $\mathbf{e}_i=\mathbf{v}_i-\mathbf{v}_p$ is the {\em edge} between points $p$ and $i$ (note that
 $\mathbf{a}_{1(2)}'=\mathbf{e}_{1(2)}$). $\boldsymbol{\nu}_{i,i+1}$ is the normal to edges $\mathbf{e}_i$ and $\mathbf{e}_{i+1}$ ($i$ is a cyclic index), and  $\hat{\mathbf{n}}=\frac{\sum_{i=1}^6\boldsymbol{\nu}_{i,i+1}A_i}{\sum_{i=1}^6A_i}$ is the area-weighted normal with $A_i=|\mathbf{e}_i\times \mathbf{e}_{i+1}|/2$ \cite{Math1,usgeo}. (See Ref.~\onlinecite{usgeo} for extensive discussion.) The mean curvature can enhance the spin-orbit coupling in carbon-based nanomaterials \cite{Ando2000,Huertas-Hernando}.

\begin{figure}[h]
\begin{center}
\includegraphics[width=0.6\textwidth]{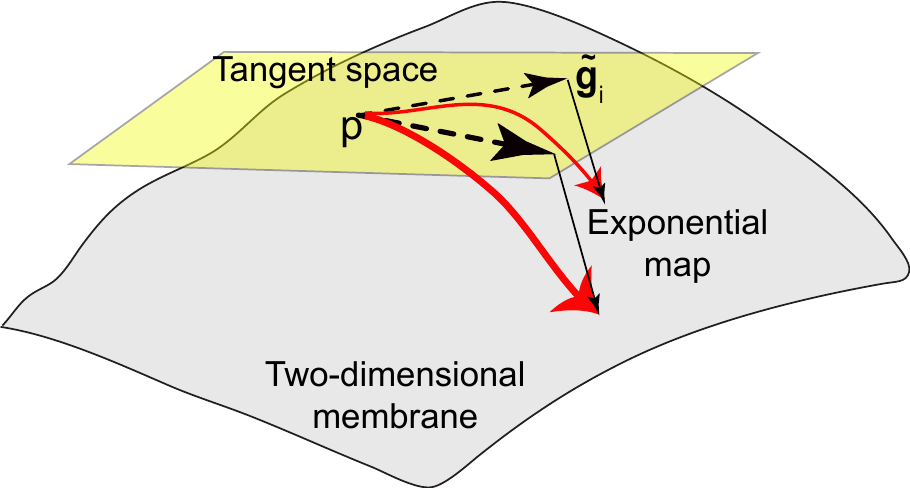}
\end{center}
\caption{In three-dimensional materials, a deformation field always belong to the bulk. In 2D conformal crystals --on the other hand-- a continuum deformation field may lie outside of the material body. Structural stability requires redefining the standard Cauchy-Born rule\cite{Landau}, so that deformations are mapped back into the two-dimensional material. Being defined onto the atomistic membrane, discrete geometry does not require such a mapping.}
\end{figure}

\begin{figure*}
\begin{center}
\includegraphics[width=0.99\textwidth]{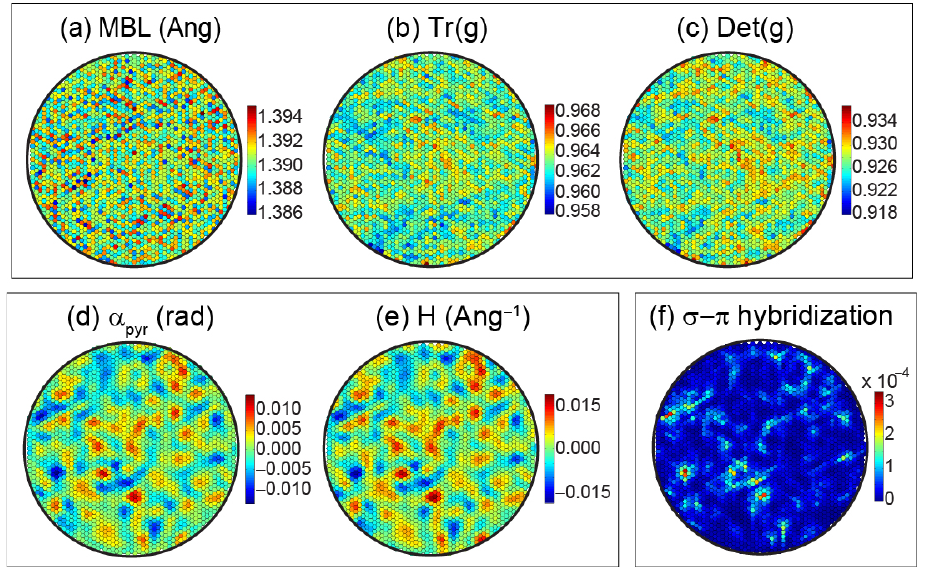}
\end{center}
\caption{Chemical measures and the local geometry of a rippled 2D crystal with hexagonal symmetry. While no direct correlation can be drawn among MBL in (a) and metric measures in (b-c), the piramidalization $\alpha_{pyr}$ in (d) is directly proportional to the mean curvature $H$ in (e); this holds for all systems studied. (f)  $\sigma-\pi$ mixing \cite{Manes} is small ($\sim 10^{-4}$).}
\end{figure*}

\subsection{Additional remarks on exponential maps}
 On the discrete geometry one never leaves the atomistic membrane because all quantities are expressed in terms of actual atomic positions. On continuum approaches to mechanics/geometry --on the other hand-- it is quite possible that tangents lie outside of the 2D manifold, and this must be remediated. The well-known Cauchy-Born rule applies to {\em space-filling} materials, but not to crystalline membranes, unless corrected with an exponential map \cite{M2,M3,M4,Ericksen,usgeo}. The acute need for exponential maps becomes yet another difference among 2D conformal crystals and bulk 3D materials. The need for exponential maps is not a technical matter, but a fundamental geometrical requirement for geodesic curves in 2D manifolds.  Exponential maps are deeply-rooted and long-known for the geometrical description of surfaces \cite{doCarmo,Lee,Kuhnel}, and must be enforced onto descriptions of nanoscale phenomena in conformal crystalline membranes.

\section{Discussion}
\subsection{Relations among chemical measures and the discrete local geometry}
Even though the relation among mean-bond lengths and metric is direct, the degree of sensitivity of MBL with respect to fluctuations on interatomic distances makes a direct correlation difficult \cite{Fasolino1,us,usSSC}. In Figures 3(a-c) we contrast MBL with  $Tr(g)$ and $det(g)$ (we will not display the determinant of the metric tensor in other figures; see Ref.~\onlinecite{usgeo} for extensive discussion). Details of the creation of the rippled structure can be found in prior work\cite{us,usSSC,usgeo} (see Methods section too). Recalling the notion that pristine graphene has equal bond lengths, Figure 3(a) indicates that atomistic fluctuations will have a bearing on the aromatic behavior of rippled samples; this concept --nor its ramifications-- has not been discussed before. Further analysis requires specific theoretical tools beyond our reach (see {\em e.g.}, Ref.~\onlinecite{FowlerSteiner}). \footnote{Here we mention that the structure was obtained at 1 Kelvin (see Methods). At this temperature the lattice constant is of the order of 1.39 \AA, but we use the more common value of 1.41 \AA{} when normalizing; the metric in Figures 3(b-c) reflects this normalization.}

 On the other hand, there exists a remarkably simple, one-to-one correlation among the pyramidalization angle and the mean curvature (sign included) for all the systems studied, as already evident from Figure 3(d-e). Put it simply, we find:
\begin{equation}\label{eq:correlation}
\alpha_{pyr}\text{ (in rads) }\simeq 1\times H \text{ (in \AA$^{-1}$).}
\end{equation}
Equation (8) is an interesting result because it embodies a commonly used angular measure for orbital hybridization and chemical reactivity with a (simple!) geometrical character. This result will hold for other hexagonal 2D crystals, and deserves additional discussion.

$\alpha_{pyr}$ is a signed quantity, as follows: Direct inspection of Figure 1(b) and Equation (3) indicates that $\alpha_{pyr}$ will be positive for a bulge, and negative for a sag. Similarly, the mean curvature $H$ in Equation (7) is a vector quantity projected onto the local normal; the relative orientation of the normal (facing ``up'' or ``down'') confers $H$ with a sign as well. (Geometrically speaking, one sees that radius of curvature changes sign for a bulg or a sag so $H$ must in fact be signed.) But the correspondence goes beyond the sign. The cross products on $H$ --Equation (7)-- confers an additional sinusoidal function, which approximates as the angle rather well up to 20 degrees (0.35 rad), which is within the range of all pyramidalization angles seen on this work. The correlation in display in Equation (8) is remarkable as it informs our intuition concerning hybridization, thus making the mean curvature a {\em direct} tool for analysis of hybridization and chemical reactivity for 2D systems with $s$ and $p$ electrons.

 Now, the coefficient $(\hat{n_1}\cdot\hat{r})(\hat{n_2}\cdot\hat{r})$ in Equation (4) approximates the amount of $\sigma-$ mixing onto $\pi-$ electrons in graphene and renormalizes the Fermi velocity in that system. Here we see that its magnitude is negligible ($10^{-4}$) for this sample of rippled graphene [Figure 3(f)]. (It may be worth mentioning here the existence of a roughly quadratic relation among $H$ and this coefficient, which is not pursued further.)

\begin{figure*}
\begin{center}
\includegraphics[width=0.9\textwidth]{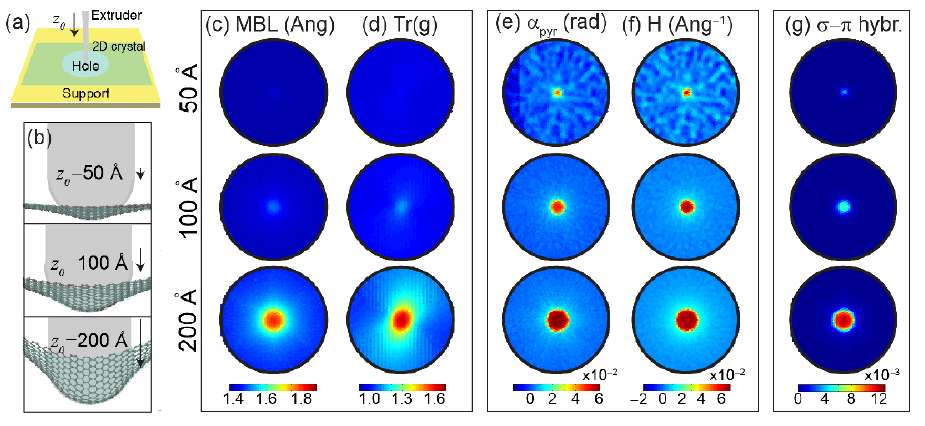}
\end{center}
\caption{Chemical measures and the local geometry of a rippled 2D crystal with hexagonal symmetry under mechanical load.}
\end{figure*}

 Let us next consider these geometrical and chemical measures when the 2D membrane in Figure 3 is under mechanical load. We worked with a large sample containing a few million atoms and extensive technical details have been given before \cite{us,usSSC,usgeo}. Let us concentrate on geometrical aspects relevant to Chemistry next.

 Figure 4(a) provides basic schematics of the mechanical indentation procedure: The 2D crystal is held fixed outside the central circular region, and a spherical extruder pushes the 2D membrane down. Data is analyzed at indentation heights of 50, 100, and 200 \AA{} [Figure 4(b)]. To highlight atomistic detail, the same area  was employed as in Figure 3. The discrete metric is formed by the lattice vectors on the deformed manifold, and it is not orthogonal. This shows on Figure 4(d), which lacks spherical symmetry\cite{usgeo}. Nevertheless, the indentation irons out the ripples seen in Figure 3 before indentation, and this helps develop a one-to-one relation among MBL and Tr(g) [Figures 4(c) and 4(d)]. Note interatomic distances as big as 1.8 \AA, well beyond the harmonic elastic regime \cite{us}.

 Interestingly, the relation among $\alpha_{pyr}$ and $H$ still remains, and $H$  [Figure 4(f)] reproduces all and every single detail from $\alpha_{pyr}$ [Figure 4(e)], even though they are obtained following very different premises (Equation 1 {\em versus} Equation 7). Even under extreme mechanical load, the coefficient $(\hat{n_1}\cdot\hat{r})(\hat{n_2}\cdot\hat{r})$ in Equation (4) [Figure 4(g)] is still smaller than a few percent; an encouraging result as it grants a sound footing to the effective theories of massless fermions in graphene in terms of $\pi-$electrons only --even under such rather strong mechanical deformation \cite{usgeo}.

\subsection{Atomistic defects and mixed (atomistic/continuum) approaches to the geometry of 2D conformal crystals}

\begin{figure}[bt]
\begin{center}
\includegraphics[width=0.7\textwidth]{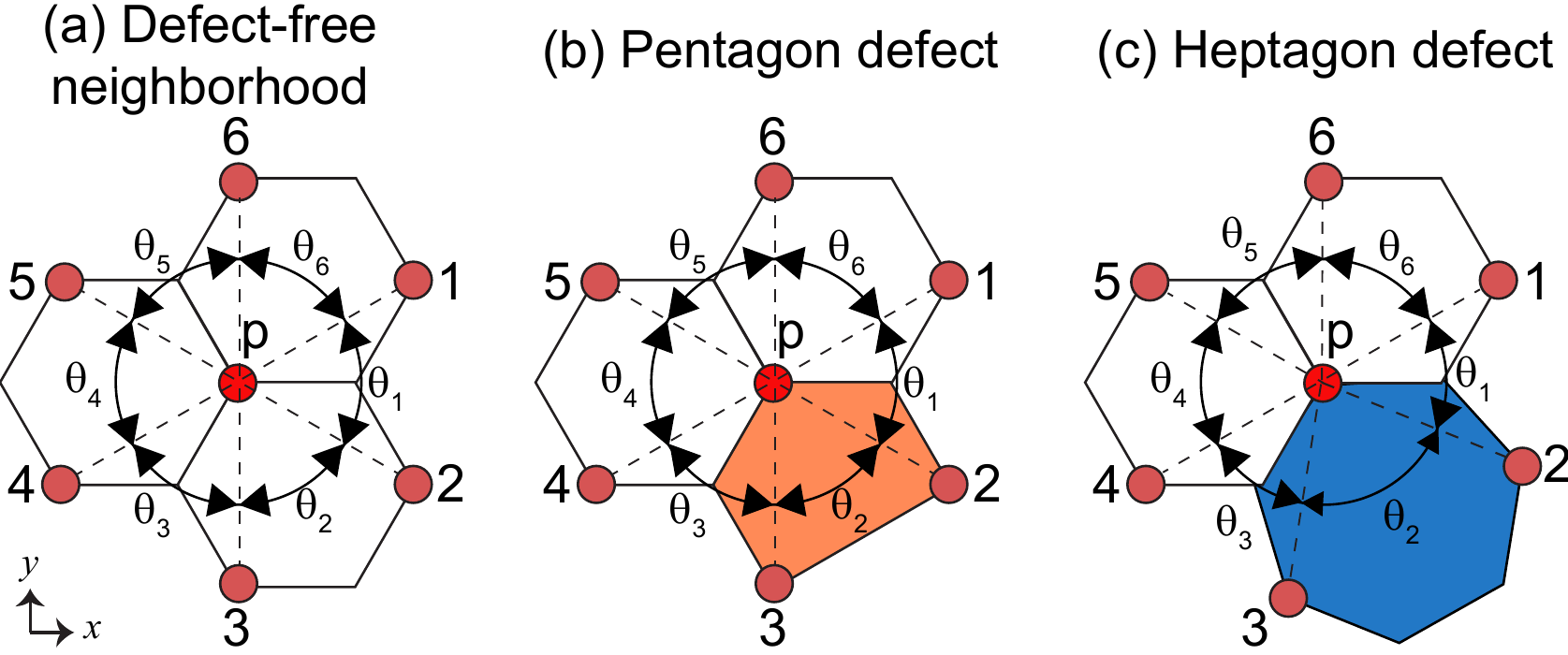}
\end{center}
\caption{The neighborhood of an atom changes upon appearance of topological defects. (a) Neighborhood with no topological defects (see Fig.~1(c)). The more common defects are: (b) a pentagon defect,  (c) a heptagon defect, or a combination of (b) and (c). Although the curvatures can still be obtained for a (b) and (c) by considering the atoms shown in red, the discrete metric breaks down, and a continuum metric is given by interpolation methods.}
\end{figure}

The pentagon (heptagon) defect in Figure 5(b) will furnish a positive (negative) Gauss curvature $K$. The presence of defects therefore leads to an infinite variety of structures. Topological defects break the crystal symmetry, and the crystalline metric [Equation (5)] will not be well-defined around atomistic defects. Hence, we develop next a metric in terms of continuous, differentiable vector fields that interpolate atomic positions.

Incidentally, this discussion is quite relevant at this moment, and it finds a unique place within the context of continuum formalism for ``strain-engineering'' of 2D crystals \cite{Ando2002,Pereira1,GuineaNatPhys2010,Vozmediano,deJuanPRB,Dejuan2011,deJuanPRL2012,deJuan2013,Kitt2013,Peeters3,Peeters4,PeetersNew1,PeetersNew2,Naumis1,Naumis2,Manes,Crommie,usold,stroscio,deheer2,us,usSSC,usgeo,strainGuinea2013,PereiraNL2013} as it presents with clarity ``black-box'' details of continuum approximations from atomistic meshes which have not been discussed up to this point. The specific approach presented here relies on the atomic lattice, it works with and without defects, and constitutes another ingredient of novelty on the present manuscript. (Care must be exercised in that atomistic defects play the fundamental and non-negligible role of (pseudo)magnetic monopole sources on the effective electron theories for carbon-based materials \cite{Vozmediano1992}.)

\begin{figure}[h]
\begin{center}
\includegraphics[width=0.9\textwidth]{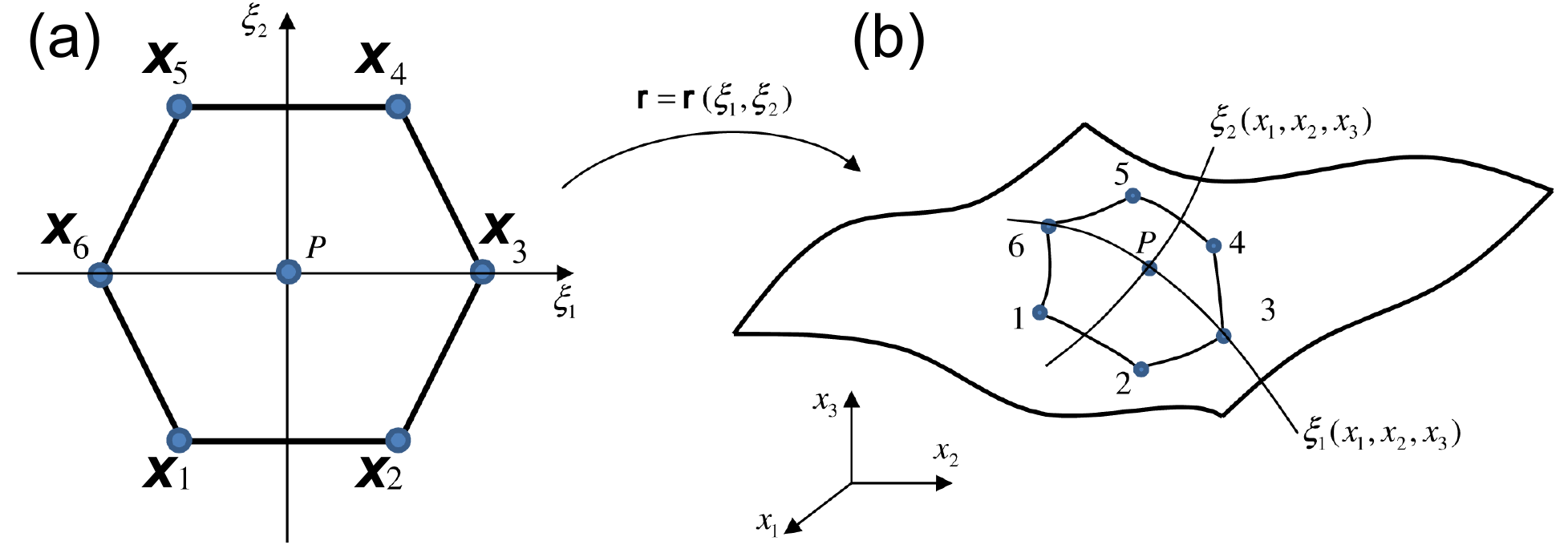}
\end{center}
\caption{(a) Orthogonal two-dimensional convective coordinate set ($\xi^1$,$\xi^2$) and a polygon. (b) The deformed polygon on an actual 2D crystal. Partial derivatives, necessary in defining the continuum metric, are carried on the convective space.}
\end{figure}

In the approach presented here all nodal points originate from atoms. This approach, appropriately called {\em conforming polygonal}, is employed nowadays for structural analysis \cite{s1}. Consider a point $p$ with coordinates ($\xi^1,\xi^2$) inside the regular polygon on the plane in Figure 6(a), and the polygon with the same number of sides formed by $n_{p}$ atoms and coordinates $\mathbf{r}_i=(x_i,y_i,z_i)$ $(i=1,...,n_{p})$ on the conformal 2D crystal (Figure 6(b)). A point with continuum coordinates $\mathbf{r}=(x,y,z)$ inside the conformal 2D crystal (Figure 6(b)) is found by interpolation:
\begin{equation}
\mathbf{r}=\sum_{i=1}^{n_{p}}\mathbf{r}_i\phi_i(\xi^1,\xi^2),
\end{equation}
and the continuum metric is expressed from vector fields in terms of the (convective) coordinates ($\xi^1,\xi^2$):
\begin{equation}
\mathbf{g}_{\alpha}=\frac{\partial\mathbf{r}}{\partial \xi^{\alpha}} \text {, and } g_{\alpha\beta}=\mathbf{g}_{\alpha}\cdot \mathbf{g}_{\beta} \text{ ($\alpha,\beta=1,2$)}.
\end{equation}
This process is carried out at each polygon in the sample, and the metric displayed in Figures 7 and 8 was evaluated at each polygon's center.
 The (Laplace) interpolating (shape) functions in Equation (9) are defined by:
\begin{equation}
\phi_i(\xi^1,\xi^2)=\frac{w_i(\xi^1,\xi^2)}{\sum_{j=1}^{n_p}w_j(\xi^1,\xi^2)},
\end{equation}
with $w_i$ a function of areas $A$ given in terms of three points defining the polygon on the convective space [Figure 6(a)]:
\begin{equation}
w_i(\xi^1,\xi^2)=\frac{A(\mathbf{x}_{i-1},\mathbf{x}_{i},\mathbf{x}_{i+1})}{A(\mathbf{x}_{i-1},\mathbf{x}_{i},(\xi^1,\xi^2))\times A(\mathbf{x}_{i},\mathbf{x}_{i+1},(\xi^1,\xi^2))}.
\end{equation}
We keep this discussion short; extensive information is given in Ref.~\onlinecite{s1}.

 We have just indicated that we will be approximating the metric to discuss surfaces with topological defects, but curvature must be dealt with additional care. Deep-ingrained perceptions from continuum geometry --such as a curvature evolving smoothly from point to point-- break down in crystalline surfaces, where {\em discrete objects} --atoms and bonds-- concentrate and {\em carry all curvature} \cite{Math1}. For this fundamental reason, we express curvature in systems with atomistic defects from the six points highlighted in Figure 5, still employing Equations (5) and (6) to extract geometrical information, thus carrying on with the discrete approach to curvature even in the presence of atomistic defects.
  We display in Figures 7 and 8 the chemical relevant measures and the local geometry for a host of interesting structures with topological defects. All relations found for 2D crystals with no defects, and discussed in Figures 3 and 4, continue to hold.

\begin{figure*}[tb]
\begin{center}
\includegraphics[width=0.99\textwidth]{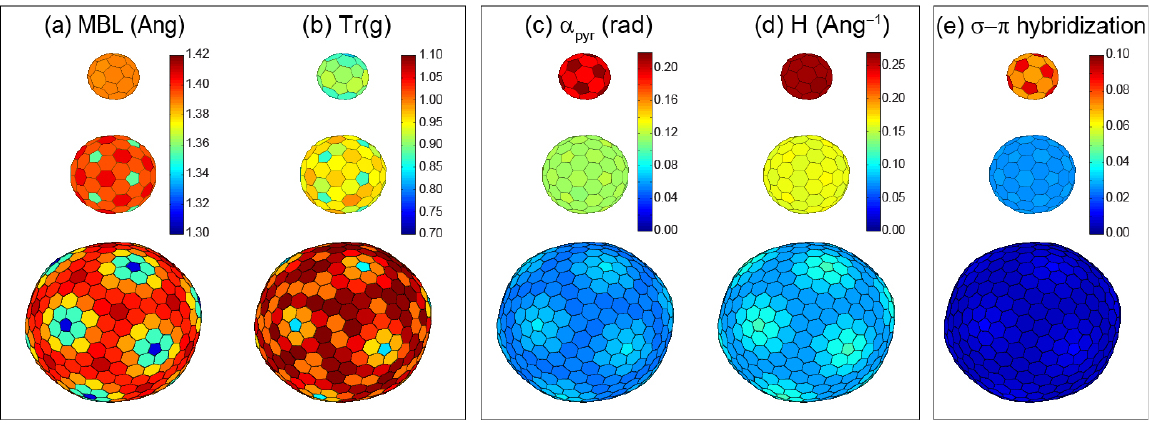}
\end{center}
\caption{Chemical and geometrical measures for three ionic crystals \cite{WalesPRB2009}.}
\end{figure*}

\begin{figure*}[tb]
\begin{center}
\includegraphics[width=0.99\textwidth]{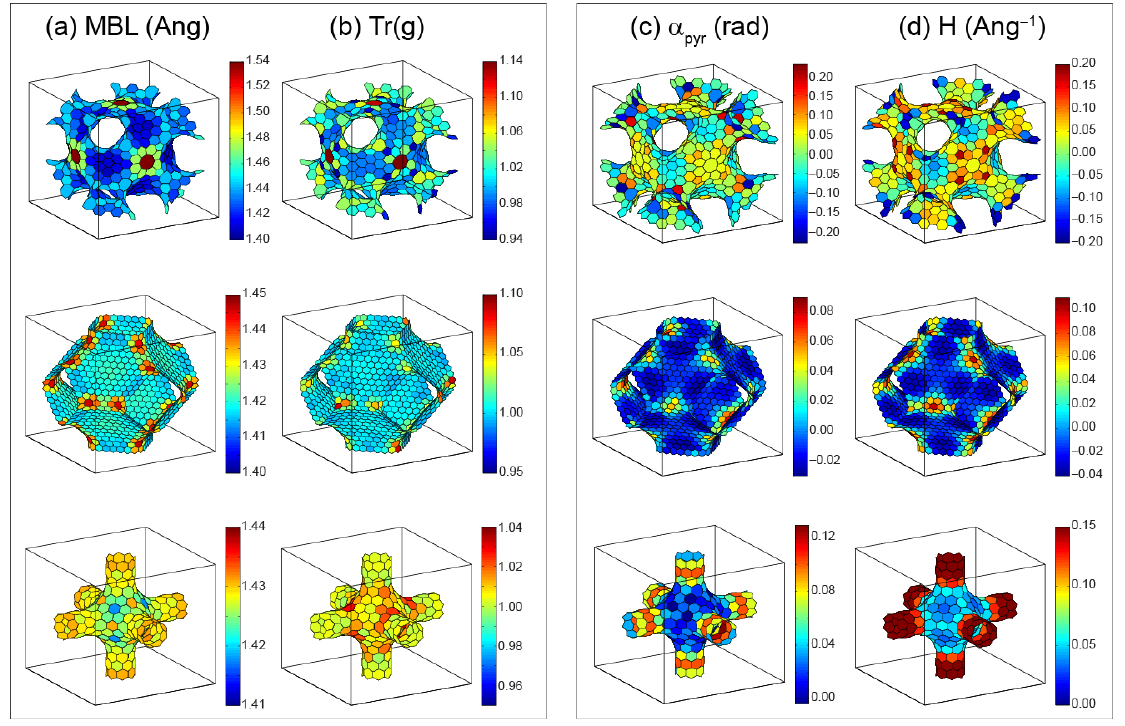}
\end{center}
\caption{Chemical and geometrical measures for three space-filling 2D crystalline materials.}

\end{figure*}

\subsection{Additional remarks on kinetics and thermodynamics}
 In considering the kinetics and thermodynamics of two-dimensional crystals, energy barriers to atomic rearrangements may also depend on shape \cite{Wales2,Wales3,WalesBook}. Thus, the geometry discussed here --in terms of atoms alone-- may prove very useful in establishing quantitative connections among shape, kinetics and thermodynamics. These connections may prove essential, for example, to study potential growth mechanisms of 2D crystals on conformal geometries.

\section{Conclusion}

 We proposed in the recent past a discrete theory for Dirac fermions on graphene, and soon found the need to explore the geometrical framework in which these theories rest. As the family of 2D atomic crystals (graphene, hexagonal boron nitride, silicene, germanene, novel binary semiconductors), and (layered) transition metal dichalcogenides and topological insulators continues to grow, establishing atom-based geometrical tools for analysis of the properties of these 2D systems (both graphene and beyond graphene) is a justified and current endeavor.

  By their dimensionality, 2D crystals can sustain large deformations, well beyond the reach of continuum elasticity theory. In addition, continuum elasticity only works in the bulk, as in 2D crystals a continuum deformation field may lie outside of the material body.

We exemplify the discrete geometry and uncover a linear relation among a chemically-relevant measure (hybridization angle) and a fundamental geometrical measure (mean curvature). In spite of the longevity of the field of atom-thin materials, such insightful identification has never been discussed before, and it may become useful for chemical insights as the field continues to evolve with ever greater vigor.

The concepts were illustrated on a number of carbon-based systems under rippling, mechanical load, on ionic crystals with spherical symmetry, and on some other crystalline structures including Schwarzites. Those concepts will translate effortlessly into other 2D crystals.

 We also discussed ways to merge continuum and discrete approaches for analysis of 2D conformal crystals. By explaining the nuts and bolts of such approaches, we decisively contribute to the discussion of continuum frameworks and atomistic approaches to uncover the properties of 2D crystals.
 Geometry from atoms is a concept relevant to Chemistry; and it will have an impact on better informing our intuition regarding the chemical properties of 2D crystals with shape, an area of research experiencing a renewed thrust.

\section{Methods}
The structures generated to study rippling --created
by line stress at the edges-- and mechanical load, were obtained from molecular dynamics \cite{LAMMPS} of a 3-million-atom (finite)
graphene square patch with no periodic boundary conditions in thermal equilibrium at 1 Kelvin. The geometrical study focused on a small patch near the geometrical center.

\begin{acknowledgement}
 We thank NSF-XSEDE {TACC's \em Stampede} (Grant TG-PHY090002) and Arkansas ({\em Razor II}) for computational support.
\end{acknowledgement}



\providecommand*\mcitethebibliography{\thebibliography}
\csname @ifundefined\endcsname{endmcitethebibliography}
  {\let\endmcitethebibliography\endthebibliography}{}

\end{document}